\documentclass[pra,floatfix,amsmath,showpacs]{revtex4}

\usepackage{graphicx}
\usepackage{amssymb}
\usepackage{mathptm,times}
\usepackage{amsfonts,amssymb,latexsym,amsmath,amscd}

\usepackage{graphicx}
\usepackage{pst-all}

\newcommand{\ket}[1]{|#1\rangle}
\newcommand{\bra}[1]{\langle#1|}

\newcommand{\bl}{\begin{itemize}}
\newcommand{\el}{\end{itemize}}

\newcommand{\tr}{\text{tr}}

\begin{document}
\title{Anyonic Quantum Walks}
\author{Gavin K. Brennen$^{1}$}
\author{Demosthenes Ellinas$^2$}
\author{Viv Kendon$^{3}$}
\author{Jiannis K. Pachos$^{3}$}
\author{Ioannis Tsohantjis$^2$}
\author{Zhenghan Wang$^4$}
\affiliation{$^1$Centre for Quantum Information Science and Security, Macquarie
University, 2109, NSW Australia}
\affiliation{$^2$ Department of Sciences, Division of Mathematics,
Technical University of Crete, GR - 73 100, Chania, Crete, Greece}
\affiliation{$^3$School of Physics and Astronomy, University of Leeds, Leeds LS2 9JT, UK}
\affiliation{$^4$ Microsoft Research, Station Q, University of California,
Santa Barbara, CA 93106, USA}
\date{\today}

\begin{abstract}

The one dimensional quantum walk of anyonic systems is presented. The
anyonic walker performs braiding operations with stationary anyons of the
same type ordered canonically on the line of the walk. Abelian as well as
non-Abelian anyons are studied and it is shown that they have very different
properties. Abelian anyonic walks demonstrate the expected quadratic quantum
speedup. Non-Abelian anyonic walks are much more subtle. The exponential
increase of the system's Hilbert space and the particular statistical
evolution of non-Abelian anyons give a variety of new behaviors. The
position distribution of the walker is related to Jones polynomials,
topological invariants of the links created by the anyonic world-lines
during the walk. Several examples such as the SU$(2)_k$ and the quantum
double models are considered that provide insight to the rich diffusion
properties of anyons.

\end{abstract}


\maketitle

\section{Introduction}

Quantum versions of random walks came to prominence in quantum information
theory through the search for efficient new quantum algorithms.  Since
classical random walks provide the techniques for some of the best
classical algorithms, it was natural to look for a quantum equivalent.
Typically, a quantum walk gives a quadratic algorithmic speed up over classical ones
\cite{shenvi02a} and for some problems an exponential speed up is possible \cite{childs02a}.  This
is a generalization of the faster spreading behavior shown by the
simplest quantum walk, a single walker on an infinite line.

Quantum walks have since found wider applications than their algorithmic
origins. In essence a quantum walk is a discrimination of a diffusion
process. For example, they has been employed in modeling transport of charge
or energy in biological systems \cite{Lloyd+Aspuru-Guzik08}, in physical
systems to show delocalization and quantum coherence
\cite{travaglione01a,dur02a,sanders02a}, as a model system
interpolating between quantum and classical behavior
\cite{kendon04a}, as well as a model amenable
to quantum simulation of its asymptotics~\cite{Ellinas}.  On the algorithmic
side they have been proved to be universal for quantum computation
\cite{childs08b,lovett09a} and they have been physically implemented, both
as a computation on a quantum computer \cite{ryan05a} and a physical walk
(e.g.~\cite{bouwmeester99a,meschede09a}). Quantum walks have been studied theoretically on many different structures: in higher dimensions~\cite{Mackay+}, under the effects of
decoherence~\cite{Kendon+Tregenna}, with Dirac spin particles~\cite{Ellinas1}, and with a
huge range of variations on the basic walk dynamics. For accessible introductions and reviews, see, for
example,~\cite{kempe03a,kendon06a,kendon06b}.

Here we focus on the question of how the quantum walk distribution can be
affected by the statistical properties of the walker.  Non-interacting
bosons in quantum walks have effectively already been shown to have
identical properties to single quantum walkers through experiments with
coherent light~\cite{bouwmeester99a,kendon04a}. Omar {\em et al.}
\cite{omar04a} studied quantum walks with two bosonic or fermionic walkers.
Entanglement in the initial state of these walkers gives a strikingly
different behavior with respect to the standard quantum walk.

More exotic statistics than the bosonic or the fermionic can be found in two-dimensional systems where
anyons can appear.  Exchanging two anyons can introduce phase factors or
unitary transformations. These evolutions are
different representations of the braiding group. Envisioning how quantum
walks can be performed with anyons one quickly realizes that the problem
takes the form of diffusion of a walker-anyon in the presence of other
regularly arranged static anyons. In particular, we embed a one dimensional quantum walker into a two dimensional medium with a canonical line ordering of the anyons from left to right. When the anyonic walker is moving among the static anyons then anyonic braiding is realized evolving the overall state of the system in a non-trivial way. Hence, by considering an anyonic walk we enrich the one particle diffusion problem with statistical properties.

Anyonic quantum walks is not only an academic curiosity, but it can be
considered as a quantum simulator modeling the diffusion of extended
objects. In order to keep track of the statistical properties of anyons it
is useful to visualize their world-lines that encode their braiding history.
When the walker moves from its initial position it braids with the straight
world-lines of the static anyons. Finding the walker at a certain position
after a number of steps requires the consideration of all the different
paths of the same length with the same initial and final positions. This
gives rise to world-line links the complexity of which increases fast with
respect to the number of the walker steps. The diffusion of extended objects
has applications in biophysics and polymer physics \cite{Nechaev}. Here, we
are interested in the quantum version of such diffusion processes.

Before turning to the anyonic quantum walks we briefly introduce the
standard quantum walk as well as the anyons and their properties. In
Sec.~\ref{walking} we present in detail the general formalism of the anyonic
quantum walk. In Sec.~\ref{abelian} we demonstrate that Abelian anyonic
quantum walks have similar asymptotic behavior as the standard quantum
walks. In Sec.~\ref{nonabelian} we consider the non-Abelian anyonic quantum
walks and we express the walker probability distribution in terms of the
Kauffman brackets. Several examples are explicitly solved for a small number
of walker steps that indicate a wide range of diffusion properties ranging
between the classical and the quantum walks. Finally, in Sec.~\ref{outlook}
we present our conclusions and we point out interesting future directions.

\subsection{Quantum walk on a line}

Quantum walks have been studied in both the continuous time
\cite{farhi98a} and discrete time \cite{ambainis01a,aharonov00a} versions.
We briefly describe the discrete time quantum walk on the infinite line,
which will be later generalized to the anyonic case. It is defined in direct
analogy with a classical random walk: there is a walker carrying a coin
which is tossed each time step and the walker moves left or right according
to the heads or tails outcome of the coin toss.

We denote the basis states for the quantum walk as an ordered pair of labels
in a ``ket'' $\ket{s}_{\rm space}\otimes\ket{j}_{\rm spin}$, where
$s\in{\mathbb Z}$ is the position and $j\in \{0,1\}$ is the spin-like state
of the coin. A unitary coin operator is used at each time step and then a
shift operation is applied to move the walker to its new positions. The
simplest coin toss is the Hadamard operator $H$, defined by its action on
the basis states $\ket{s,j}$ as
\begin{eqnarray}
H \mid s, 0 \rangle &= \frac{1}{\sqrt{2}}(\mid s, 0 \rangle + \mid s, 1
\rangle) \nonumber \\
H \mid s, 1 \rangle &= \frac{1}{\sqrt{2}}(\mid s, 0 \rangle - \mid s, 1
\rangle),
\label{eqn:hadamard}
\end{eqnarray}
and the shift operation $T$ acts on the basis states as
\begin{eqnarray}
T \mid s, 0 \rangle &= \mid s-1, 0 \rangle \nonumber \\
T \mid s, 1 \rangle &= \mid s+1, 0 \rangle.
\label{eqn:shift}
\end{eqnarray}
A single step of the quantum walk consists of $W=TH$ applied to the quantum
walker plus spin. After performing $t$ steps of the quantum walk, where the
walker is initially in state $\mid \Phi(0)\rangle$, we obtain
the final state
\begin{equation}
\mid\Phi(t)\rangle = W^t\mid\Phi(0)\rangle.
\label{eqn:qw}
\end{equation}
Ultimately we are interested in the probability distribution $P(s,t)$ of the spatial location, $s$, of the walker at time $t$ that is given by
\[
(P(1,t),P(2,t),...)=\text{diag}\left(\tr_{\rm
spin}\big[\ket{\Phi(t)}\bra{\Phi(t)}\big] \right).
\]

As the walk progresses, quantum interference occurs whenever there is more
than one possible path of $t$ steps that leads to the same position. This
interference is both constructive and destructive, which causes some
probabilities to be amplified or decreased at each timestep. This leads to
the different behavior compared to its classical counterpart. In the latter
the position of a walker, following a classical random walk on a line,
spreads out in a binomial distribution about its starting point.  In Refs.~\cite{ambainis01a}, Ambainis {\it et al.} proved that the quantum walk on a
line spreads in $O(t^2)$ compared to a classical random walk which spreads
in $O(t)$, where the spreading is
being measured as the second moment about a single initial starting point.

\subsection{Anyons and their properties}

It is commonly accepted that point-like particles, elementary or not, come
in two species, bosons or fermions. This statistical label is determined by
the behavior of their wave function when two identical non-interacting
particles are exchanged. For bosons the wave function is unchanged while for
fermions it acquires a phase of $\pi$, i.e., changes sign. These are the
only observed statistical behaviors for particles that exist in our three
dimensional world. If one restricts to two dimensions there are more
possibilities in the particle statistics. In this case, when two particles
are exchanged their wave function can be evolved by an arbitrary phase
factor or even a unitary operator that creates superpositions in an internal
space of the particles. These particles, named anyons by Frank Wilczek, can
appear as effective quasiparticles, such as vortices, in two dimensional
many body systems. They are manifested in the Fractional Quantum Hall Effect
(FQHE)~\cite{Laughlin,Camino} and they are expected to appear in p-wave
superconductors~\cite{Rice,Tewari}, topological insulators~\cite{Fu,Teo} and
a variety of lattice models~\cite{Kitaev,Pachos1,Pachos2,Pachos3,Levin}.

Anyons can be created from the vacuum in pairs, they can be braided around
each other and they can be fused together. While the pair creation results
in a particle and its antiparticle much in the same way as usual particles,
braiding operations and fusion of anyons are rather unique. This is due to
the non-local degrees of freedom carried between anyons that give rise to
their exotic statistics. One can think of it like an internal `spin'. When
two anyons are braided then their internal `spin' state either acquires a
phase factor, $e^{i\phi}$ (Abelian anyons) or it gets evolved by a unitary
matrix, $b$ (non-Abelian anyons). The fusion corresponds to the tensor
product of these `spins' with a restriction on the maximum possible `spin'
value. When two anyons, $a$ and $b$, are fused then more than one outcomes
are possible depending on their internal state. Symbolically we have $
a\times b=c+d+... $.

\begin{figure}[h]
\begin{center}
\includegraphics[scale=0.6]{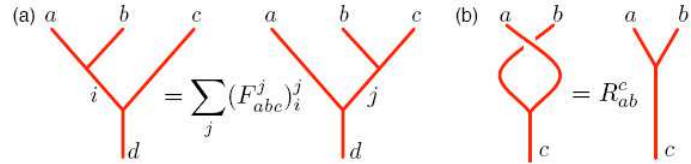}
\caption{\label{fig:FR} The (a) $F$ and (b) $R$ moves dictated
from the fusion and braiding properties of anyons.}
\end{center}
\end{figure}
Three anyons $a$, $b$ and $c$ can be fused to a final anyon $d$ in two
different ways. Either one can fuse $a$ with $b$ and its result $i$ with $c$
or one can fuse $b$ with $c$ and its result $j$ with $a$. These two
different processes correspond to two different bases of the collective
state of the anyons.  The different ways of combining particles
$a,b,c$ to yield $d$ is given by the recoupling formula:
$\ket{(ab)c\rightarrow
d;x}=\sum_{x'}(F_{abc}^d)_x^{x'}\ket{(a(bc)\rightarrow d;x'}$ as seen in
Fig.~\ref{fig:FR}(a). The braiding of two anyons $a$, $b$ with fusion
channel $c$ is described by $R_{ab}^c$ as seen in Fig.~\ref{fig:FR}(b). To
obtain the braiding result of two anyons that do not have a direct fusion
channel (see for example anyons $b$ and $c$ on the left hand side of
Fig.~\ref{fig:FR}(a)) one can use the $F$ matrix to obtain their braiding
element, $F^{-1}RF$. Anyons also carry a quantum dimension $d$. This
accounts for the dimension of the internal space of the anyons and provides
the scaling of the total Hilbert space. For example, $n$ anyons with quantum
dimension $d$ have a Hilbert space with dimension $d^n$. Abelian anyons have
$d=1$ and non-Abelian ones have necessarily $d>1$. Note that, in contrast to
usual spin, $d$ can be an irrational number. For more details on anyons and
on topological quantum computation see, e.g.~\cite{BrennenPachos} and
references therein.

\section{Walking anyons}
\label{walking}

\begin{figure}[ht]
\begin{center}
\includegraphics[scale=0.6]{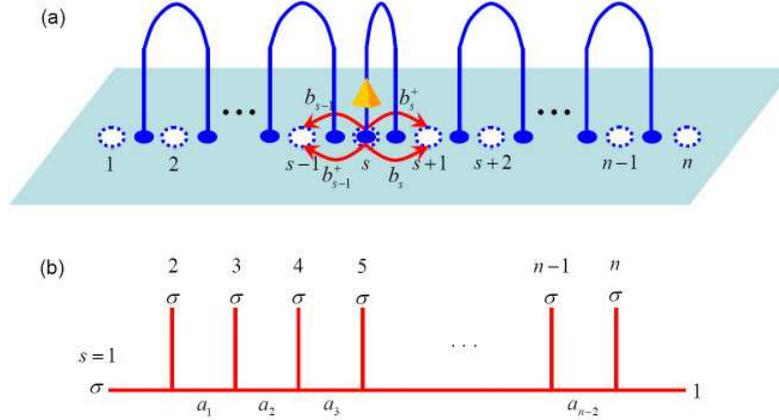}
\caption{\label{fig:2} The anyonic quantum walk.
(a)  A quantum walk of the anyonic walker with a four state die attached.
It walks along one of four braids depending on the state of the die on sites
labelled by $s=1$ to $s=n$ (white circles) positioned in between stationary anyons (blue circles). (b)  A basis state
$\ket{\Psi(a_1,a_2,\ldots, a_{n-2})}$ of the non-local fusion space
representing the sequential topological charge measurements
 of $n$ type $\sigma$ anyons with total charge zero
(vacuum, also denoted as $1$).   }
\end{center}
\end{figure}

In order to reveal the statistics of anyons in a quantum walk, we consider a
single anyonic walker braiding around others in fixed positions. The anyons
involved are all of the same type $\sigma$.  The walker has an attached die
degree of freedom which gives it distinguishability.  However this
distinguishability does not negate the action of braiding on the non-local
statistical degrees of the system. The anyons are placed in a canonical
order in the plane with position labelled by an integer $s$, as seen in
Fig.~\ref{fig:2}(a). For an anyon at position $s$, other anyons with
positions $s'<s$ $(s'>s)$ will be said to be on the left (right). We
restrict to $n\bmod 4=2\times$odd in order to have an odd number of pairs
such that  the initial state with the marked anyonic walker and its partner
located at the middle will have an equal number of pairs to their right and
left.  This facilitates the construction, if desired, of a symmetric
evolution.

Basis states are labelled by $\ket{\Psi(a_1,a_2,\ldots a_{n-2})}$ according
to the fusion outcomes as indicated by Fig.~\ref{fig:2}(b). The state with
nearest neighbor vacuum pairs depicted in Fig.~\ref{fig:2}(a) is
\[
\ket{\Psi_0}=\ket{\Psi(1,\sigma,1,\sigma,\ldots,
1,\sigma)},
\]
where $1$ is the vacuum particle, as can be verified by considering the
consistent fusion ordering. If desired one could modify the protocol to give
a symmetric distribution for the walker by changing the initial state to
provide for a symmetric spatial distribution for the anyons.  Taking the
vacuum partner of the walker and winding counterclockwise halfway around all
its neighbors, the initial fusion state $\ket{\Psi_0}=b_{n}b_{n-1}\cdots
b_{(n+3)/2}\ket{\Psi(1,\sigma,1,\sigma,\ldots, 1,\sigma)}$ provides for the
walker anyon to have the same number of vacuum pairs on its left and right.

The Hilbert space of our system is
\begin{equation}
\mathcal{H}(n)=\mathcal{H}_{\rm anyons}(n)\otimes \mathcal{H}_{\rm die}(n).
\end{equation}
The die contains both a discrete spatial index indicating its position at
the location of the anyons on the plane and $4$ internal spin states:
$\mathcal{H}_{\rm die}(n)=\mathcal{H}_{\rm space}(n)\otimes \mathcal{H}_{\rm
spin}$, where $\mathcal{H}_{\rm space}={\rm span}_{\mathbb
C}\{\ket{s}\}_{s=1}^{n}$ and $\mathcal{H}_{\rm spin}={\rm span}_{\mathbb
C}\{\ket{j}\}_{j=0}^{3}$. The anyonic Hilbert space is further decomposed as
$\mathcal{H}_{\rm anyons}(n)=\mathcal{H}_{\rm local}(n)\otimes
\mathcal{H}_{\rm fusion}(n)$  the first factor describing degrees of freedom
local to each anyon and the later factor the non-local or fusion degrees of
freedom.  We restrict to a contractible surface so that the homological
degrees of freedom are trivial.   Furthermore, the local degrees of freedom,
while measurable, will be conserved during this walk and will be ignored.

Fusion degrees of freedom are determined by the fusion rules $a_i\times
a_j=\sum_k N_{a_ia_j}^{a_k}a_k$, where $N_{ab}^c\in~\mathbb{N}$ counts the
number of ways to combine anyons of type $a$ and $b$ to obtain $c$.  The
corresponding Hilbert state space of $n$ $\sigma$ anyons with total charge
zero is $\mathcal{H}_{\rm fusion}(n)={\rm
span}_{\mathbb{C}}\{\ket{\Psi(\{a_j\}_{j=1}^{n-2})}\}$ with dimension ${\rm
dim}(\mathcal{H}_{\rm fusion}(n))=\sum_{a_1,a_2,\ldots,a_{n-2}}N_{\sigma
\sigma}^{a_1}N_{a_1\sigma}^{a_2}N_{a_2\sigma}^{a_3}\cdots
N_{a_{n-2}\sigma}^{1}$.  We will restrict to cases where the fusion spaces
are one dimensional, i.e. $N_{ab}^c< 2$.  Braiding
relations are determined by the $F$ and $R$ matrices. The action of the generators
$\{b_s\}_{s=1}^{n-1}$ of the braid group $\mathcal{B}_{n}$ on
$\mathcal{H}_{\rm fusion}(n)$  is given by the amplitudes
\begin{equation}
\bra{\Psi(\{a'_j\})}b_k\ket{\Psi(\{a_j\})}=\left\{
\begin{array}{lc}  \displaystyle\prod_{m=k}^{n-2}\delta_{a'_m,a_m}
R_{\sigma\sigma}^{a_1}, & k=1 \\
\displaystyle\prod_{m=k}^{n-2}\delta_{a'_m,a_m} \sum_x
([F_{\sigma\sigma\sigma}^{a_k}]^{-1})_x^{a'_{k-1}}R_{\sigma\sigma}^x
(F_{\sigma\sigma\sigma}^{a_k})_{a_{k-1}}^x, & k=2 \\
\displaystyle\prod_{\ell=1}^{k-2}\delta_{a'_{\ell},a_{\ell}}
\displaystyle\prod_{m=k}^{n-2}\delta_{a'_m,a_m}
\sum_x ([F_{a_{k-2}\sigma\sigma}^{a_k}]^{-1})_x^{a'_{k-1}}R_{\sigma\sigma}^x
(F_{a_{k-2}\sigma\sigma}^{a_k})_{a_{k-1}}^x,& 2<k< n-1\\
\displaystyle\prod_{m=k}^{n-2}\delta_{a'_m,a_m} R_{\sigma\sigma}^{a_{n-3}} , & k=n-1
 \end{array}\right.,
\nonumber
\end{equation}
where the formula $b=F^{-1}RF$ is employed when the walker does not have
direct fusion channels with the rest of the anyons.

The initial state of our system is
\[
\ket{\Phi(0)}=\ket{\Psi_0}_{\rm fusion}\ket{s_0=
\frac{n}{2}}_{\rm space}\ket{\psi}_{\rm spin}.
\]
The quantum walk algorithm consists of an iterative sequence of operations
$W=TU$ where $U$ is the die tossing operation that acts trivially on the
position of the walker and $T$ is the conditional braiding operator. We can
introduce a virtual tensor product structure to the spin states:
$\ket{0}_{\rm spin}=\ket{0}_{\rm x}\ket{0}_{\rm y},\ket{1}_{\rm
spin}=\ket{0}_{\rm x}\ket{1}_{\rm y},\ket{2}_{\rm spin}=\ket{1}_{\rm
x}\ket{0}_{\rm y},\ket{3}_{\rm spin}=\ket{1}_{\rm x}\ket{1}_{\rm y}$ where
$\ket{0}_{\rm y}(\ket{1}_{\rm y})$ are modes which are behind (in front of)
their neighbors, and $\ket{0}_{\rm x}(\ket{1}_{\rm x})$ are modes moving
left or right. For the die toss we pick (in the product basis)
\[
U=\frac{1}{\sqrt{2}}\left(\begin{array}{cc}1 & i \\i & 1\end{array}\right)
\otimes \frac{1}{\sqrt{2}}\left(\begin{array}{cc}1 & i \\i & 1
\end{array}\right)=e^{i\frac{\pi}{4}(X_x+X_y)},
\]
which can be interpreted as a pair of beam splitters on the two axes. Each component of $U$ is a symmetrized version of the Hadamard operation of Eqn. (\ref{eqn:hadamard}).
Of course a continuum of tossing operations could be considered but this decomposition into a virtual tensor product has advantages as described below.  The conditional braiding operation $T$ has the following action
\[
T\ket{\Psi}_{\rm fusion}\ket{s}_{\rm space}\ket{\kappa}_{\rm x}\ket{\gamma}_{\rm y}=b_{s+\kappa-1}^{(-1)^{\kappa\oplus_2\gamma}}\ket{\Psi}_{\rm fusion}\ket{s+2\kappa-1}_{\rm space}\ket{\kappa}_{\rm x}\ket{\gamma}_{\rm y}
\]
over the spatial range $0<s<n+1$.

It is convenient to introduce the unnormalized spinor
\[
 \ket{\vec {\Phi}(s,t)}=(_{\rm spin}\bra{0}\Phi(s,t)\rangle,_{\rm spin}\bra{1}\Phi(s,t)\rangle,_{\rm spin}\bra{2}\Phi(s,t)\rangle,_{\rm spin}\bra{3}\Phi(s,t)\rangle)^T,
 \]
so that an arbitrary state at time $t$ is
\[
\ket{\Phi(t)}=\sum_{s=1}^{n+1}\ket{\vec {\Phi}(s,t)}.
\]
In order to avoid boundary conditions on the walk we study the behavior for $t<n/2$.

\section{Abelian anyonic quantum walk}
\label{abelian}

\subsection{Variance of Abelian anyonic walks}

Employing the method of Brun {\em et al.}~\cite{Brun1,Brun2,Brun3} one can
analytically evaluate the variance of the distributions
corresponding to the Abelian anyonic walks by determining the moments of the
resulting distribution. For simplicity we relabel the positions so that the initial walker position in the middle of the anyonic chain is given by $s_0=0$, i.e. $\ket{\Phi(0)} =
\ket{s=0}_{\rm space}\otimes\ket{\psi}_{\rm spin}$. In Fourier components we have
$$
\ket{s=0} = \int_{-\pi}^\pi {dk \over 2\pi} \ket{k}.
$$
In the Fourier transformed basis the evolution operator $W$ acts as
$W(\ket{k} \otimes \ket{\psi}) = \ket{k} \otimes M_k\ket{\psi}$ so
after $t$ steps the state evolves to
$$
W^t\ket{\Phi(0)} = \int_{-\pi}^\pi {dk \over 2\pi} \ket{k}\otimes
(M_k)^t\ket{\psi}.
$$

One can write a closed form for the moments of the
resulting distribution. By employing the expansion of the identity in the
position $s$ basis, $1\!\!\!1 = \sum _s \ket{s}\bra{s}$ and $\langle
s|k\rangle =e^{-iks}$ we obtain
\begin{eqnarray}
\langle s^m\rangle_t &=& {1 \over (2\pi)^2} \sum_s s^m \int dk \int dk'
e^{-is(k-k')}\bra{\Phi(0)} (M_k^\dagger)^t(M_{k'})^t\ket{\Phi(0)}
\nonumber\\
&=& {i^m\over 2\pi} \int dk\int dk'\delta^{(m)}(k-k')
\bra{\Phi(0)} (M_k^\dagger)^t(M_{k'})^t\ket{\Phi(0)}.
\end{eqnarray}
Now we can integrate by parts the derivatives that act on the
$\delta$-function. For the case of $m=1$ we obtain
$$
\langle s\rangle_t = {i \over 2\pi} \int dk\bra{\Phi(0)}
(M_k^\dagger)^t{d(M_k)^t \over dk} \ket{\Phi(0)}.
$$
We have that $ {d M_k/ dk} = -i(P_R-P_L) M_k$, where $P_{R,L}$ are the
projector operators for moving the walker right or left with $P_R+P_L=1$.
Finally, we obtain
\begin{equation}
\langle s\rangle_t = {1 \over 2\pi} \sum_{j=1}^t \int dk\bra{\Phi(0)}
(M_k^\dagger)^j (P_R-P_L)(M_k)^j \ket{\Phi(0)}.
\label{firstmoment}
\end{equation}
Similarly for $m=2$ we have
\begin{equation}
\langle s^2\rangle_t = {1 \over 2\pi} \sum_{j=1}^t \sum_{j'=1}^t \int dk
\bra{\Phi(0)} (M_k^\dagger)^j (P_R-P_L)(M_k)^{j-j'} (P_R-P_L)(M_k)^{j'}
\ket{\Phi(0)}.
\label{secondmoment}
\end{equation}
One can further employ the eigenstates, $\ket{\lambda_{l}(k)}$, and
eigenvalues, $\lambda_{l}(k)$, of $M_k$ ($l$ runs through the eigenstates) to
explicitly evaluate the moments. Let us expand the initial state in this
basis
$
\ket{\Phi(0)} =\sum_l c_{l}(k)\ket{\lambda_{l}(k)}.
$ After $t$ steps the state becomes $ (M_k)^t\ket{\Phi(0)} = \sum_l
e^{i\lambda_{l}(k) t} c_{l}(k)
\ket{\lambda_{l}(k)}.
$
With this in mind the first moment becomes
$$
\langle s\rangle_t = t - {1 \over \pi} \int dk \sum_{l,l'} c_{l}^*(k)c_{l'}(k)
\bra{\lambda_{l}(k)}P_L\ket{\lambda_{l'}(k)}\sum_{j=1}^t e^{i[\lambda
_{l'}(k)-\lambda_{l}(k)]j}.
$$
If the matrix $M_k$ is non-degenerate then most of the terms in $\langle
s\rangle_t$ will be oscillatory and they will average to zero over time. The
non-zero contributions comes from the diagonal terms
\begin{equation}
\langle s\rangle_t = \Big(1-{1\over \pi}\int dk \sum_l |c_{l}(k)|^2
\bra{\lambda_{l}(k)}P_L\ket{\lambda_{l}(k)}\Big)t+\text{Oscillatory terms}.
\label{x}
\end{equation}
Similarly for the second moment we have
\begin{equation}
\langle s^2\rangle_t = \Big(1 -{2\over \pi} \int dk \sum_l |c_{l}(k)|^2
\bra{\lambda_{l}(k)}P_L\ket{\lambda_{l}(k)}\bra{\lambda_{l}(k)}P_R\ket{\lambda_{l}(k)}\Big)t^2+
\text{Oscillatory terms and lower $t$ orders}.
\label{xx}
\end{equation}

For the Abelian anyonic walker,
\begin{equation}
M_k=e^{ik}M_++e^{-ik}M_-=e^{-ikZ_x}e^{i\phi Z_x\otimes Z_y}
e^{i\frac{\pi}{4}(X_x+X_y)}.
\label{MAbl}
\end{equation}
The eigenvalues of $M_k$ come in conjugate pairs
\begin{equation}
\lambda_{1}(k)=e^{-i\beta_-(k)},\quad \lambda_{2}(k)=e^{i\beta_-(k)}
\quad \lambda_{3}(k)=e^{-i\beta_+(k)},\quad \lambda_{4}(k)=e^{i\beta_+(k)},
 \end{equation}
where
$
\beta_{\pm}(k)=\cos^{-1}\Bigg[\frac{1}{2}
\Big(\cos(k)\cos(\phi)\pm \sqrt{(\cos^2(\phi)-2)\cos^2(k)+2}\Big)\Bigg]$.

\begin{figure}[h,t]
\begin{center}
\includegraphics[scale=0.9]{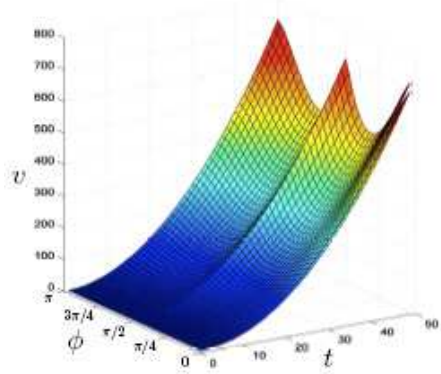}
\caption{\label{fig:Abelian1} The variance, $v=\langle
s^2\rangle-\langle s\rangle^2$, of the Abelian quantum walk as a function of
the number of steps, $t$, and the statistical angle, $\phi$. The two plots
shown are the analytical results (bottom surface), based on Eqn. (\ref{x})
and (\ref{xx}), and the numerical ones (top surface) giving almost identical
results. A quadratic behavior of the variance is witnessed as expected from
the standard quantum case.}
\end{center}
\end{figure}

The eigenvectors are
\begin{eqnarray}
&&\ket{\lambda_1(k)}=\ket{v(\beta_-)},\quad
\ket{\lambda_2(k)}=-iY\otimes Z\overline{\ket{\lambda_1(k)}},
\nonumber\\
&&
\ket{\lambda_3(k)}=\ket{v(\beta_+)},\quad
\ket{\lambda_4(k)}=-iY\otimes Z\overline{\ket{\lambda_3(k)}},
\end{eqnarray}
where
\begin{eqnarray}
\ket{v(\beta)}&=&\frac{1}{\sqrt{\mathcal{N}}}\Big[(e^{-2i(k+\beta)}+1+
e^{-2i\beta}-e^{i(-k+\phi-\beta)}+e^{2i(\phi-\beta)}-2
e^{i(-k+\phi-3\beta)}-e^{i(k+\phi-\beta)})\ket{0}_x\ket{0}_y\nonumber\\
&&\quad\quad+i(e^{-i(\phi+k+\beta)}-e^{-2i(k+\beta)}-1+e^{-i(k-\phi+\beta)})\ket{0}_x\ket{1}_y\nonumber \\
&&\quad\quad+i(e^{-i(\phi+k+\beta)}-1-e^{-2i\beta}+e^{i(k+\phi-\beta)})\ket{1}_x\ket{0}_y
+(e^{2i(\phi-\beta)}-1)\ket{1}_x\ket{1}_y\Big],
\end{eqnarray}
with $1/\sqrt{\mathcal{N}}$ the normalization factor. Employing the
eigenvectors and eigenvalues of $M_k$ and using Eqs. (\ref{firstmoment}),
(\ref{secondmoment}) the variance, $v=\langle s^2\rangle-\langle
s\rangle^2$, can be evaluated, as seen in Fig.~\ref{fig:Abelian1}.

\subsection{Results}

It is rather straightforward to perform a numerical simulation of the
Abelian quantum walk on the line for relatively large number of steps. This
gives the exact distribution of the walker and one can obtain the
corresponding variance, $v={\langle s^2\rangle-\langle s\rangle^2}$, as a
function of the number of steps, $t$, and the statistical angle, $\phi$, of
the anyons. The numerically obtained variance and the analytical results of
the previous subsection are depicted in Fig.~\ref{fig:Abelian1}, appearing
to be almost identical, with a small difference due to the neglect of terms
in the analytic calculations. We observe the expected quadratic behavior
that corresponds to the standard quantum case. There is a weak dependence on
the anyonic statistical angle $\phi$. This is due to the form of the coin
evolution that includes the term $e^{i\phi Z_y\otimes Z_x}$ (see Eq.
(\ref{MAbl})). For the case where $\phi$ is a multiple of $\pi/2$ this term
has a tensor product structure that splits the system into the tensor
product of two quantum walks, each one with a two dimensional coin. Thus, we
expect to obtain the same variance as the standard quantum walk with a
single two dimensional coin. Note that these cases include the quantum walks
of bosons ($\phi=0$) and fermions ($\phi =\pi$). For generic values of
$\phi$ the walk entangles the two coins. As a result the interference effect
that gives the speed up of the quantum evolution gets diluted over the four
dimensional space of the two coins and the slope of the variance is
decreased. This is in agreement with previous studies, where quantum walks
with multiple coins were considered~\cite{Brun1,Brun2}.

\section{Non-Abelian anyonic quantum walk}
\label{nonabelian}

In this section we would like to perform a quantum walk with non-Abelian
anyons. For simplicity we slightly change the quantum walk. Now the walker
moves to the left and to the right by always braiding in the anti-clockwise
sense with its neighboring anyons.  This can be achieved by employing only
one coin (which has two degrees of freedom). One can show that for a given
non-Abelian anyonic model the braidwords that result from this quantum walk
have the same computational power as arbitrary braidwords. Initially, note
that having static $n$ anyons and braiding one moving anyon clockwise or
anticlockwise around them can realize the full computational power of the
anyonic model~\cite{Simon}. Restricting only to anti-clockwise braidings
does not restrict the possible obtained unitaries. Indeed, for the finite
braiding groups we consider here one can show that there exists an integer
$k$ such that for a generic braid element $b_i$ it is $b_i^k=1\Rightarrow
b_i^{k-1}=b_i^{-1}$ (see~\cite{Freedman}). Thus, braidings equivalent to
clockwise moves can be produces given enough anti-clockwise ones. Notably,
fixing the chirality of the braiding means that Abelian sectors of any
physical theory will not contribute to the walker's distribution since all
walking paths accumulate the same overall phase.  Hence the single coin
protocol probes the truly non-Abelian statistics.

Consider $t$ steps of the quantum walker and, for brevity, define
$$
B^t_{\vec{a}}\equiv\prod_{r=0}^{t-3}b_{s_t+2(a_1+...a_{t-r})-(t-r)}
b_{s_t+2a_1+a_2-2}b_{s_t+a_1-1}.
$$
The evolution of the system can be given
by the unitary operator
\begin{equation}
V^t = \sum_{s_t \in [t+1,n-t+1]} \sum_{\vec{a}\in\{0,1\}}\prod_{r=1}^t
P_{a_r} U \otimes B^t_{\vec{a}}\otimes
\ket{s_t+2(a_1+...+a_t)-t}\bra{s_t}.
\end{equation}

Starting from the walker in position $m$ and the coin qubit in state
$\ket{\psi}$ one obtains after $t$ steps the state
\begin{equation}
\ket{\Psi(t)} = V^t\ket{\psi}\ket{\alpha}\ket{m} =
\sum_{\vec{a}}(\prod_{r=1}^t P_{a_r}U) \ket{\psi}
B_{\vec{a}}^t\ket{\alpha}\ket{\bar n_{\vec{a}}^t},
\end{equation}
where $\vec{a}$ is a $t$ dimensional vector with $r$-th component 0 or 1
depending if the walker is moving, respectively, left or right at the $r$-th
step, $\bar n_{\vec{a}}^t =m+2(a_1+...+a_t)-t$ and the initial fusion state
of the anyons is $\ket{\alpha} = \ket{0101...01} $ as in~\cite{Aharonov}.
One can write the corresponding density matrix $\rho^t
=\ket{\Psi(t)}\bra{\Psi(t)}$ as
\begin{equation}
\rho(t) =\sum_{\vec{a},\vec{a}'}(\prod_{r=1}^t P_{a_r}U) \ket{\psi}\bra{\psi}
(\prod_{r=1}^t P_{a'_r}U) B_{\vec{a}}^t\ket{\alpha}\bra{\alpha}
{B_{\vec{a}'}^t}^\dagger \ket{\bar n_{\vec{a}}^t}\bra{\bar n_{\vec{a}'}^t}.
\label{rhooft}
\end{equation}
Importantly, the coin, walker position and anyonic Hilbert spaces
can be traced out independently. For brevity let us set
$$
{\mathcal{U}}_{\vec{a}\vec{a}'}^t =(\prod_{r=1}^t P_{a_r}U)
\ket{\psi}\bra{\psi} (\prod_{r=1}^t P_{a'_r}U)
$$
and
$$
\mathcal{Y}_{\vec{a}\vec{a}'}^t =B_{\vec{a}}^t\ket{\alpha}\bra{\alpha}
{B_{\vec{a}'}^t}^\dagger.
$$
We would like to obtain the position distribution of the walker regardless
of the coin and anyonic states. For that we need to trace the coin states
and the fusion space of the anyons resulting in
$$
\tr_\text{coin} \tr_\text{anyon} \rho(t) =\sum_{\vec{a},\vec{a}'}
\tr {\mathcal{U}}_{\vec{a}\vec{a}'}^t
\tr \mathcal{Y}_{\vec{a}\vec{a}'}^t
\ket{\bar n_{\vec{a}}^t}\bra{\bar n_{\vec{a}'}^t}.
$$

Due to the choice of coin manipulations one needs to have paths $\vec{a}$
and $\vec{a}'$ that end up at the same coin state for the trace to be
non-trivial. To simplify the presentation we consider the composite loop
$\vec{a} \vec{a}'$ corresponding to paths $\vec{a}$ and $ \vec{a}'$ that
start at the same point and end at the same point after the same number of
steps. Thus, they are contributing to the diagonal part of
$\tr_\text{coin}\tr_\text{anyon}\rho(t)$.

For a Hadamard coin flip operation $U=H$ (see Eq. (\ref{eqn:hadamard})) one
can easily evaluate that
$$
\tr {\mathcal{U}}_{\vec{a}\vec{a}'}^t = {1 \over 2^n} (-1)^{z},
$$
where $z$ is the number of all successive pairs of $1$'s in the loop
$\vec{a}\vec{a}'$. Hence, successive zeros and ones in the loop
$\vec{a}\vec{a}'$ can be compressed. While the coin trace is rather
straightforward care needs to be taken when one takes the trace over the
anyonic degrees of freedom. It can be shown~\cite{Aharonov} that the trace
of unitary representations of the braiding group corresponds to the Markov
trace of the braid graphs described above. As initial condition we adopt the
anyons to be created pairwise from the vacuum. Hence, the trace of the
braidings gives the Kauffman bracket of the Plat traced braidings.

\begin{figure}[h]
\begin{center}
\includegraphics[scale=0.35]{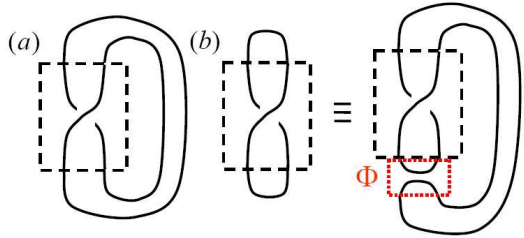}
\caption{\label{fig:MarkPlatTrace} The Markov (a) and Plat (b) tracing of a braid.
In (b) the equivalence of the Plat and the Markov traces is depicted, which
is established with the help of the Temperley-Lieb algebra element
$\Phi$~\cite{Aharonov} that has a simple diagrammatic representation.}
\end{center}
\end{figure}

  \begin{figure}[h]
\begin{center}
\includegraphics[scale=0.55]{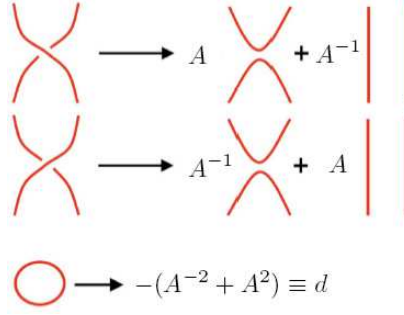}
\caption{\label{fig:kauffbrack}  Method to recursively calculate the
Kauffman bracket $\langle L\rangle(A)$ of an unoriented link $L$.  All
crossing in a projection of the link can be smoothed using the Skein
relations and components which are unknots contribute a multiplicative
factor of $d$.  Normalization is chosen so that $\langle O \rangle=1$.}
\end{center}
\end{figure}

In~\cite{Aharonov} it has been shown that
$\ket{\alpha}\bra{\alpha}=\Phi_1\Phi_3...\Phi_{n-1}/d^{n/2}$. Intuitively,
this means that the density matrix of the initial anyonic state, given by
the pair creation of particles from the vacuum, corresponds to the product
of Temperley-Lieb algebra elements $\Phi_i$ as depicted in
Fig.~\ref{fig:MarkPlatTrace}. Thus one has
$$
\tr(B^t_{\vec{a}}\ket{\alpha}\bra{\alpha} {B_{\vec{a}'}^t}^\dagger
) = {1\over
d^{n/2-1}} \langle ({B_{\vec{a}'}^t}^\dagger
 B^t_{\vec{a}})^\text{Plat}\rangle,
$$
where $B_{\vec{a}'}^\dagger B_{\vec{a}}$ have been Plat traced and then the
corresponding Kauffman bracket has been evaluated.   The Kauffman
bracket~\cite{Kauffman} of a link $\langle L\rangle(A)$ is a Laurent
polynomial in the complex variable $A$ and can be computed by smoothing each
crossing according to the Skein relation shown in Fig. \ref{fig:kauffbrack}.
Up to a phase factor they are equal to the Jones polynomials~\cite{Jones},
topological invariants that facilitate the distinction between inequivalent
links. Kauffman brackets have the following property, $\langle
L\text{O}\rangle = d\langle L\rangle$ for $L$ being a general non-empty link
and $\text{O}$ being the trivial link. Also $\langle
\text{O}\rangle=1$. Hence $\langle \text{OO}... \text{O}\rangle = d^{m-1}$
where the bracket is over $m$ trivial links that correspond to $n=2m$
anyons. In particular, when $B_{\vec{a}}$ and $B_{\vec{a}'}$ are trivial
(corresponding to trivial statistics or involving trivial anyonic braidings)
then $\tr(B^t_{\vec{a}}\ket{\alpha}\bra{\alpha} {B_{\vec{a}'}^t}^\dagger
)=1$, i.e. we obtain the usual quantum walk, with no anyonic effect
appearing.

\subsection{The SU$(2)$ level $k$ anyonic model}

While the main structures presented here are applicable to general
non-Abelian models we present, for concreteness, analytic calculations for
the SU$(2)_k$ topological models. These have
$A=ie^{i\pi/2(k+2)}$ so their quantum dimension is $d=
2\cos(\frac{\pi}{k+2})$ where $k$ is an integer greater or equal to $2$. All
the braiding and fusion matrices of these models are well
known~\cite{BrennenPachos}. In the following we explicitly calculate the
probability distribution in the position of the walker for $t=1$, $2$, $3$,
$4$ and $10$ number of walker steps.

Let us consider the $t=1$ case. For that walk we use two pairs of anyons
canonically ordered on a line and the walker is the third one from the left.
Then we have
$$
\sum_{\vec{a},\vec{a}'} \rho_{\vec{a},\vec{a}'}(t=1)|_{\text{diag}} =
\ket{2}\bra{2}\tr {\mathcal{U}}_{00}^1 \tr \mathcal{Y}_{00}^1+
\ket{4}\bra{4}\tr {\mathcal{U}}_{11}^1 \tr \mathcal{Y}_{11}^1.
$$
But
$$
\mathcal{U}_{00}^1=\mathcal{U}_{11}^1 = {1 \over 2}
,\,\,\text{and}\,\,
\tr \mathcal{Y}_{00}^1=\tr \mathcal{Y}_{11}^1 = {1 \over d}
\langle \text{OO}\rangle=1,
$$
so
$$
\sum_{\vec{a},\vec{a}'} \rho_{\vec{a},\vec{a}'}(t=1)|_{\text{diag}} =
{1\over 2} {1 \over d}\langle
\text{OO}\rangle(\ket{2}\bra{2}+\ket{4}\bra{4}) = {1 \over
2}(\ket{2}\bra{2}+\ket{4}\bra{4}).
$$
Hence, after one step the walker has equal probability ($1/2$) to be at the
two neighboring sites for all types of anyons and no anyonic effects appear.

Let us consider the $t=2$ case.  Here we use three pairs of anyons, where
the walker is initially located at position $s_0=3$.  One can evaluate
$$
\tr \mathcal{U}_{\vec{a}\vec{a}'}^2 ={1 \over 4}
$$
for all relevant pairs $\vec{a}\vec{a}'$, i.e. $(0,0)(0,0)$, $(1,1)(1,1)$,
$(0,1)(0,1)$ and $(1,0)(1,0)$ while it is zero otherwise. For the trace of
braidings we have
$$
\tr\mathcal{Y}_{(0,0)(0,0)}^2 = {1 \over d^2}\langle (b_1
b_2)^\dagger b_1b_2\rangle ={1 \over d^2}\langle \text{OOO}\rangle=1
$$
and similarly
$$
\tr\mathcal{Y}_{(1,1)(1,1)}^2 = \tr\mathcal{Y}_{(0,1)(0,1)}^2 =
\tr\mathcal{Y}_{(1,0)(1,0)}^2 = 1.
$$
Hence, we finally have
$$
\sum_{\vec{a},\vec{a}'} \rho_{\vec{a},\vec{a}'}(t=2)|_{\text{diag}} =
{1 \over 4}(\ket{1}\bra{1}+\ket{5}\bra{5})+{1 \over 2}\ket{3}\bra{3}.
$$
Again the anyonic character is not apparent at $t=2$ either.

\begin{figure}[h]
\begin{center}
\includegraphics[scale=0.45]{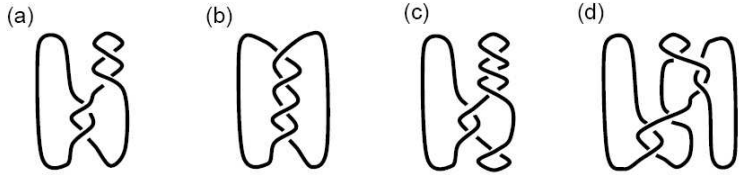}
\caption{\label{fig:t3knot} The links that contribute to the $t=3$ (a) and
$t=4$ (a-d) anyonic quantum walks.}
\end{center}
\end{figure}

Let as consider the probability distribution for $t=3$ where we use $5$
pairs of anyons. The paths $\vec{a}\vec{a}'$ that start at the common
initial position $s_0=5$ and end at the same final position having at the
same time the same final coin state are given by $(000)(000)$, $(001)(001)$,
$(010)(010)$, $(100)(100)$, $(011)(011)$, $(101)(101)$, $(110)(110)$,
$(111),(111)$, $(010)(100)$, $(100)(010)$, $(011)(101)$, and $(101)(011)$.
One can see that for all of the paths we have $\tr
\mathcal{U}_{\vec{a}\vec{a}'}^3 ={1 / 8}$ apart from the last two $(011)(101)$, and
$(101)(011)$ for which $\tr\mathcal{U}_{\vec{a}\vec{a}'}^3 =-{1 / 8}$. Also
for all of the paths we have $\tr\mathcal{Y}_{\vec{a}\vec{a}'}^3 =1$ apart
from the last four for which it is
$$
\tr ({B^3_{(010)}}^\dagger\ket{\alpha}\bra{\alpha} B^3_{(100)})+
\tr ({B^3_{(100)}}^\dagger\ket{\alpha}\bra{\alpha} B^3_{(010)}) =
{1 \over d} (\langle L\rangle+\langle L\rangle^*)={2 \cos ({2\pi \over k+2})
\cos ({3\pi\over k+2}) \over \cos ({\pi\over k+2})}
$$
and
$$
\tr ({B^3_{(011)}}^\dagger \ket{\alpha}\bra{\alpha} B^3_{(101)})+
\tr ({B^3_{(101)}}^\dagger \ket{\alpha}\bra{\alpha}B^3_{(011)}) =
{1 \over d} (\langle L\rangle+\langle
L\rangle^*)={2 \cos ({2\pi \over k+2})
\cos ({3\pi\over k+2}) \over \cos ({\pi\over k+2})},
$$
where $L$ is the non-trivial link given in Fig.~\ref{fig:t3knot}(a). Hence,
the position distribution is of the form
$$
\sum_{\vec{a},\vec{a}'} \rho_{\vec{a},\vec{a}'}(t=3)|_{\text{diag}} =
{1 \over 8}\ket{2}\bra{2}+\Big({3 \over 8} +{1\over 4}{\cos ({2\pi\over
k+2})
\cos ({3\pi\over k+2}) \over \cos ({\pi\over k+2})}\Big) \ket{4}\bra{4}+
\Big({3 \over 8} -{1\over 4}{\cos ({2\pi\over k+2}) \cos ({3\pi\over k+2})
\over \cos ({\pi\over k+2})}\Big)
\ket{6}\bra{6} +{1\over 8}\ket{8}\bra{8}.
$$
The trace of this reduced density matrix is $1$ as expected. Nevertheless,
it acquires nontrivial contributions from the anyonic nature of the walker
and its environment.

For the $t=4$ case we have
$$
\sum_{\vec{a},\vec{a}'} \rho_{\vec{a},\vec{a}'}(t=4)|_{\text{diag}} =
{1 \over 16}\ket{1}\bra{1}+ \frac{1}{8} \left(-3 \cos \left(\frac{2 \pi
}{k+2}\right)+3 \cos \left(\frac{4 \pi }{k+2}\right)+5\right)\ket{3}\bra{3}+
$$
$$
\frac{1}{32} \left(3 \cos \left(\frac{2 \pi }{k+2}\right)-\cos \left(\frac{4 \pi }{k+2}\right)-3 \cos \left(\frac{6 \pi }{k+2}\right)+5\right) \sec ^2\left(\frac{\pi }{k+2}\right)\ket{5}\bra{5}+
$$
$$
\frac{1}{16} \left(2 \cos \left(\frac{2 \pi }{k+2}\right)-\cos \left(\frac{4 \pi }{k+2}\right)+1\right) \sec ^2\left(\frac{\pi }{k+2}\right)
\ket{7}\bra{7}+{1\over 16}\ket{9}\bra{9},
$$
where the contributing links are given in Fig.~\ref{fig:t3knot}(a-d).

\begin{figure}[h]
\begin{center}
(a)\includegraphics[scale=0.55]{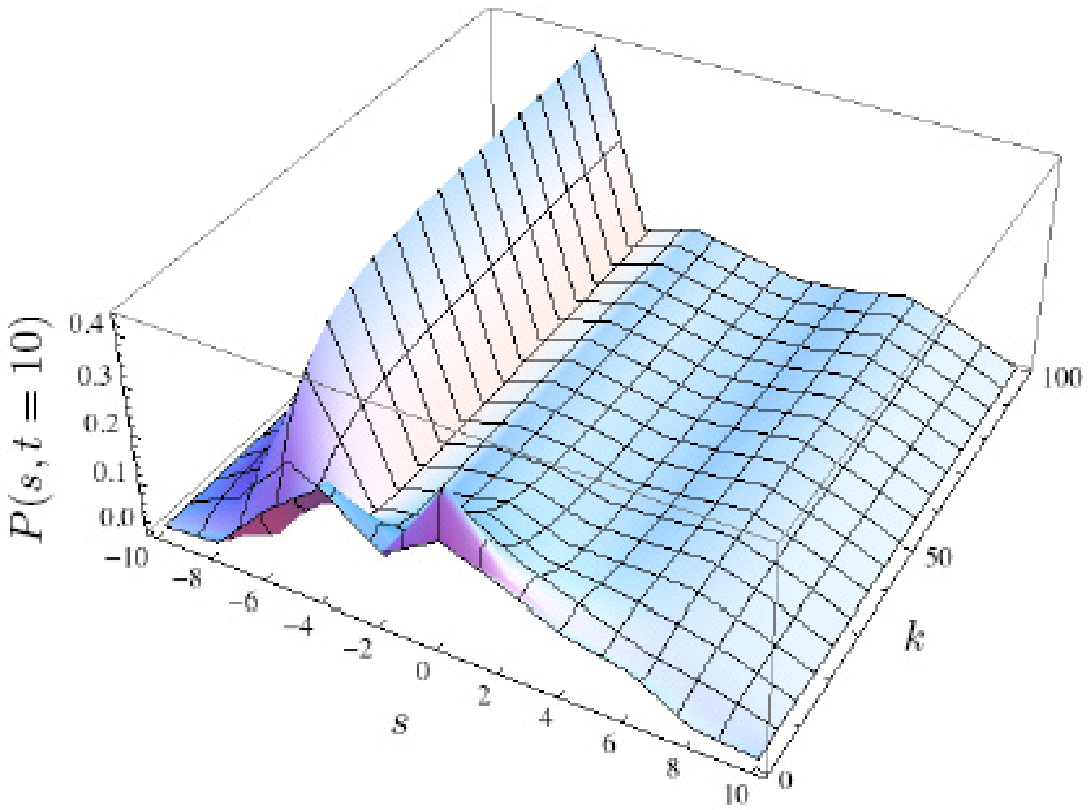}
(b)\includegraphics[scale=0.6]{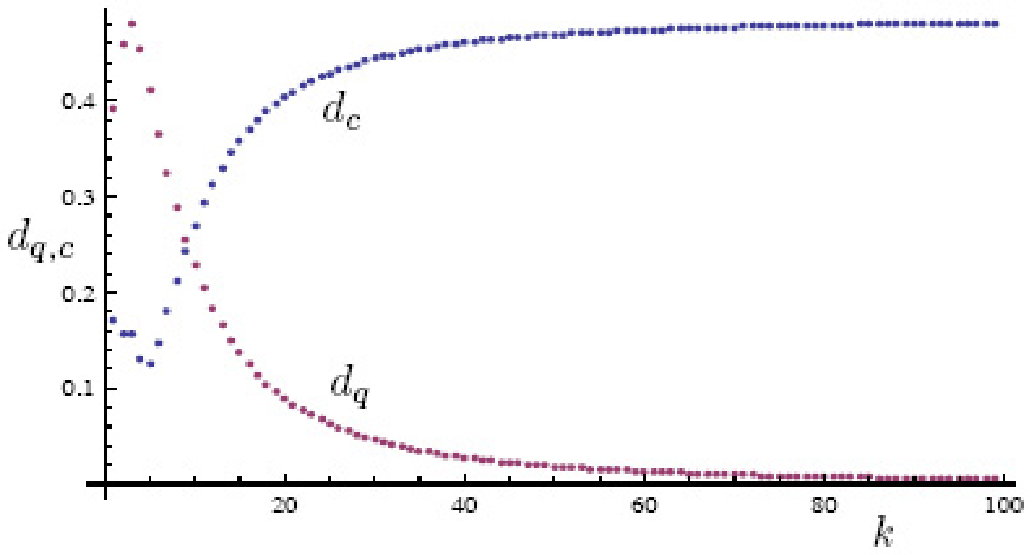}
\caption{\label{fig:distri} (a) Distribution of the quantum walker, $P(s,t=10)$,
for a $t=10$ step walk as a function of position $s$ and the type of anyonic
model indexed by $k$. The origin has been shifted so that the starting point
is $s_0=0$. (b) The distance $d_q$ and $d_c$ between the distribution
$P(s,t=10)$ and the corresponding quantum $P_q(s,10)$ and classical
$P_c(s,10)$, respectively. Clearly, for small $k$ the dispersion is closer
to the classical one and for large $k$ it is closer to the quantum one.}
\end{center}
\end{figure}

We have obtained analytic results for the distribution of the walker for a
$t=10$ step walk. This distribution is depicted in Fig.~\ref{fig:distri}(a)
as a function of the position of the walker and the level $k$ of the
SU$(2)_k$ theory. We notice that for large $k$ the distribution, $P(s,t)$,
approaches an asymptotic value. This is true for any $t$ we checked. In
order to determine the behavior of the walker distribution we compare it to
the classical and quantum walks. For that we introduce the distance between
the anyonic distribution $P(s,10)$ and the corresponding standard quantum or
classical ones $P_{q,c}(s,10)$ as
$d_{q,c}=\sqrt{\sum_s|P(s,10)-P_{q,c}(s,10)|^2}$. From
Fig.~\ref{fig:distri}(b) we see that $d_q$ goes to zero for large $k$. For
example, for $k>70$ it is $d_q<0.01$. Hence, for large $k$ the anyonic walk
distributions are very close to ones obtained from the corresponding quantum
walks. This is in agreement with the intuition that for $k\rightarrow\infty$
the anyons behave as spin-$1/2$ fermions ($\phi=\pi$ of the Abelian case),
which reproduces the standard quantum walk distribution. On the other hand
the distance to the classical random walk, $d_c$, is closer to zero for
small $k$, while in this region $d_q$ becomes maximal. From
Fig.~\ref{fig:distri}(b) we see that the minimum of $d_c=0.126$ is obtained
for $k=6$. Clearly, walks with much larger number of steps is needed in
order to conclusively determine the $t$ asymptotic behavior of this anyonic
model.

It is instructive to consider the decomposition of the anyonic Hilbert space
in terms of qubits. As a specific example consider the case of SU$(2)_2$
anyons with labels $1,\sigma,\psi$ representing particles with spin
$0,\frac{1}{2},1$ respectively.  The non-trivial fusion rules are
\[
\sigma\times\sigma=1+\psi,\quad \sigma\times \psi=\sigma,\quad \psi\times\psi=1,
\]
and the relevant non-trivial $F$ and $R$ matrices are
\[
F_{\sigma\sigma\sigma}^{\sigma}=\frac{1}{\sqrt{2}}
\left(\begin{array}{cc}1 & 1 \\1 & -1\end{array}\right),
\quad R_{\sigma\sigma}=\left(\begin{array}{cc}1 & 0 \\0 &
i\end{array}\right),
\]
expressed in the basis $\{1,\psi\}$. For the total charge zero
superselection sector, the fusion rules require that the only physical
fusion basis states are those with $a_{2k}=\sigma$ for
$k=[1,\frac{n}{2}-1]\cap \mathbb{Z}$. This then gives a fusion space
isomorphic to $m=\frac{n}{2}-1$ qubits. The generators of the braid group
$\mathcal{B}_n$ assume a particularly simple representation in
$\mathcal{H}_{\rm fusion}(n) \simeq (\mathbb{C}^2)^{\otimes m}$
\begin{eqnarray}
B_1&=&R\otimes_{j=2}^m {\bf 1}_2,\quad B_2=B\otimes_{j=2}^m {\bf 1}_2,\quad
B_3=A\otimes_{j=3}^m {\bf 1}_2 \nonumber\\
B_{2k}& =&\otimes_{j=1}^{k-1}{\bf 1}_2\otimes B\otimes_{j=k+1}^m {\bf 1}_2,
\quad B_{2k+1} =\otimes_{j=1}^{k-1}{\bf 1}_2\otimes A\otimes_{j=k+2}^m {\bf 1}_2;\quad 1<k<m \nonumber\\
B_{n-3}&=&\otimes_{j=1}^{m-2} {\bf 1}_2\otimes A,\quad
B_{n-2}=\otimes_{j=1}^{m-1} {\bf 1}_2\otimes B,
\quad B_{n-1}=\otimes_{j=1}^{m-1} {\bf 1}_2\otimes R,
\label{su2rep}
\end{eqnarray}
where
\begin{eqnarray}
R&=&
\left(\begin{array}{cc}1 & 0 \\0 & i\end{array}\right),
\quad
B=[F_{\sigma\sigma\sigma}^{\sigma}]^{-1}RF_{\sigma\sigma\sigma}^{\sigma}=
\frac{1}{\sqrt{2}}
\left(\begin{array}{cc}e^{i\frac{\pi}{4}} & e^{-i\frac{\pi}{4}}  \\
e^{-i\frac{\pi}{4}}  & e^{i\frac{\pi}{4}}
\end{array}\right),
\nonumber\\
A_{a_i}^{a_{i+2}}&=&\sum_x
[(F_{a_i\sigma\sigma}^{a_{i+2}})^{-1}]_x^{\sigma}R^x
[F_{a_i\sigma\sigma}^{a_{i+2}}]_{\sigma}^{x}\Rightarrow
A=\left(\begin{array}{cccc}1 & 0 & 0 & 0 \\0 & i & 0 & 0 \\0 & 0 & i & 0 \\0
& 0 & 0 & 1\end{array}\right).
\nonumber
\end{eqnarray}


\subsection{The $D(G)$ quantum double model}

So far we have described quantum walks with non-Abelian anyons corresponding
to irreducible representations (irreps) of the quantum group $SU(2)_{k}$.
Yet there are other physical theories with anyonic excitations that are
described by finite dimensional irreps of quasi-triangular Hopf algebras. As
with quantum groups it is possible to find representations of the braid
group induced from these algebras which then allow us to compute the
entanglement of the fusion degrees of freedom with the walker's spatial
degree of freedom. First we write down a general link polynomial
$L(z,\overline{z})$ which is a Laurent polynomial in a complex parameter $z$
and its conjugate $\overline{z}$ for oriented links. We demand that
$L(z,\overline{z})$ is an invariant under ambient isotopy. Writing the $n$
component link $L=(B)^{\mathrm{Markov} }$ as the Markov trace over a braid
word $B=\prod_{k}b_{p_{k}}^{m_{k}}\in
\mathcal{B}_{n}$, with $p_{k}\in \{1,2,\ldots ,n-1\}$, $m_{i}\in \mathbb{Z}$
, we have \cite{res,zhanggouldbracken}
\begin{equation*}
L(z,\overline{z})=(z\overline{z})^{-n/2}\Big(\frac{\overline{z}}{z}\Big)
^{e(B)/2}\varphi (B),
\end{equation*}%
where $e(B)=\sum_{k}m_{k}$ (this is the same as the writhe when we orient
all components in the same way). Here the quantity $\varphi (B)$ is the
Markov trace which satisfies the following properties
\begin{equation*}
\begin{array}{lll}
(\mathrm{I})\quad  & \varphi (BB^{\prime })= & \varphi (B^{\prime }B)\quad
\forall B,B^{\prime }\in \mathcal{B}_{n} \\
(\mathrm{II})\quad  & \varphi (Bb_{n})= & z\varphi (B)\quad \forall B\in
\mathcal{B}_{n-1}\subset \mathcal{B}_{n}, \\
& \varphi (Bb_{n}^{-1})= & \overline{z}\varphi (B)\quad \forall B\in
\mathcal{B}_{n-1}\subset \mathcal{B}_{n},
\end{array}
\end{equation*}
where $z=\varphi (b_{n})$, $\overline{z}=\varphi (b_{n}^{-1})$. As a
consequence of these properties, the link invariant satisfies
\begin{equation*}
\begin{array}{lll}
L(BB^{\prime })^\text{Markov} & = & L(B^{\prime })^\text{Markov}\quad
\forall
B,B^{\prime }\in \mathcal{B}_{n}, \\
L(Bb_{n}^{\pm 1})^\text{Markov} & = & L(B)^\text{Markov}\quad \forall B\in
\mathcal{
B}_{n-1}\subset \mathcal{B}_{n}.
\end{array}
\end{equation*}
Using the additivity of the writhe number it can been shown that for the
link invariant $L_{n}$ constructed from $\varphi _{n}$ we have that, $
L_{n}(\hat{\theta}_{1}\#\hat{\theta}_{2})=L_{i}(\hat{\theta}_{1})L_{n-
i}(\hat{\theta}_{2}) $, where $L_{i}(.)$ and $L_{n-i}(.)$ are respectively
constructed from $\varphi _{i}(.)$ and $\varphi _{n-i}(.)$.

There is a procedure to construct the braid group representation of
quasi-triangular Hopf algebras but we focus here on a large class given by
the quantum double of a finite group $D(G)$. The irreps of $D(G)$ are
labelled by the pair $(k,\alpha )$, where $k$ labels a conjugacy class $%
\mathcal{C}_{k}$ of $G$ and $\alpha $ labels an irrep, with dimension $%
|\alpha |$, of the centralizer $Z_{k}$ of a representative element $g_{k}\in
C_{k}$. The quantum dimension of these irreps (our anyons) is $d[k,\alpha ]=|%
\mathcal{C}_{k}||\alpha |$. From the representation theory of quantum
doubles it is shown \cite{gould,TG} that for the invariant above
\begin{equation}
L(z,\overline{z})=d[k,\alpha ]^{n}\langle g_{k}\rangle _{\alpha
}^{-e(B)}\varphi (B),\quad B\in \mathcal{B}_{n}  \label{lin}
\end{equation}%
with
\begin{equation*}
\varphi (B)  =  \frac{\text{tr}(B)}{(d[k,\alpha ])^{n+1}},\quad
z  =  \frac{\langle g_{k}\rangle _{\alpha }}{d[k,\alpha ]},\quad z=\frac{
\langle g_{k}^{-1}\rangle _{\alpha }}{d[k,\alpha ]},\quad
\langle g_{k}^{\pm 1}\rangle _{\alpha }=\frac{\chi _{\alpha }^{k}(g_{k}^{\pm
1})}{|\alpha |},
\end{equation*}
and $\chi _{\alpha }^{k}$ is the character in the representation $\alpha $.

Consider as an example the symmetric group $G=S_{N}$, $N\geq 5$, and $%
C_{2}\subseteq S_{N}$ \ the conjugacy class of transpositions of $S_{N}$ ,
of order $\left\vert C_{2}\right\vert =\frac{1}{2}N(N-1)$ with chosen
representative $g_{2}=(N-1\;N)$, and let $Z_{2}=S_{N-2}\times S_{2}$ be the
centralizer subgroup of $g_{2}$ where $S_{2}=\{I,\;g_{2}\}$. Let $\alpha =1$
labelling the trivial irrep of $Z_{2}$, from which the $(k,\alpha )=(2,1)$
irrep of $D(S_{N})$ with dimension $d[2,1]=\frac{1}{2}N(N-1)$ is induced
(see \cite{TG}). For this irrep $\langle g_{k=2}\rangle _{\alpha =1}=1$ and
(\ref{lin}) becomes $L(B)^\text{Markov}=d[2,1]^{n}\varphi (B).$

Consider the braid group on $n+1$ strands, $\mathcal{B}_{n}$, and a
canonical word $B=\left( b_{i_{1}}\right) ^{m_{1}}\left( b_{i_{2}}\right)
^{m_{2}}...\left( b_{i_{n}}\right) ^{m_{n}}\in \mathcal{B}_{n}$. If
$(i_{1},i_{2},...,i_{n})$ is an arbitrary permutation of ($1,2,...,n$) and
$m_{i}\in\mathbb{Z} $, the link polynomial for the trace closure of these
words reads
\begin{equation}
L(B)^\text{Markov}=\prod_{i=1}^{n}\left\{ 1+\frac{2}{3}(N-2)\left[ 1+2\cos (\frac{%
2m_{i}\pi }{3})\right] +\frac{1}{4}(N-2)(N-3)\left[ 1+(-1)^{m_{i}}\right]
\right\} .  \label{link}
\end{equation}
By virtue of (\ref{rhooft}), and by taking for coin initial state
$|\psi\rangle ,$ we have for the diagonal elements of the walker's density
matrix the following expression

\begin{eqnarray}
\sum_{\vec{a},\vec{a}^{\prime }}\rho _{\vec{a},\vec{a}^{\prime }}(t)|_{\text{%
diag}} &=&\sum_{j=1}^{t}\sum_{\mu _{j}=0}^{j}\sum_{k=1}^{t}\sum_{\nu
_{k}=0}^{k}\delta _{\mu _{t},\nu
_{t}}\delta _{\mu _{1},\nu _{1}}\mathcal{U}_{\overrightarrow{\mu },%
\overrightarrow{\nu }}  \notag \\
&&\varphi \big((\prod_{r=1}^{t}(\delta _{\mu _{r},\mu _{r-1}}+\delta _{\mu
_{r},\mu _{r-1}+1})b_{s+\mu _{r}+\mu _{r-1}-r})(\prod_{r=1}^{t}(\delta _{\nu
_{r-1},\nu _{r}}+\delta _{\nu
_{r-1},\nu _{r}-1})b_{s+v_{r}+v_{r-1}-r})^{\dagger }\big)  \notag \\
&&\left\vert \frac{n+1}{2}+2\mu _{t}-t\right\rangle \left\langle \frac{n+1}{2%
}+2\mu _{t}-t\right\vert,
\end{eqnarray}%
where ${\mathcal{U}}_{\mu ,\nu }=\text{tr}\left( \prod_{r=0}^{t}(P_{\mu
_{r+1}-\mu _{r}}U)\ket{\psi}\bra{\psi}\left( \prod_{r=0}^{t}(P_{\nu
_{r+1}-\nu _{r}}U)\right) ^{\dagger }\right)$. Here we take
$\ket{\psi}\bra{\psi}=\ket{0}\bra{0}$ and $\ \rho _\text{walker }=\left\vert
\frac{n+1}{2}\right\rangle \left\langle \frac{n+1}{2}
\right\vert$ as the initial coin and walker density matrices,
respectively. As we now employed the Markov trace rather than the Plat one
the obtained probability distribution corresponds to the modified quantum
walker algorithm where we start with $n$ anyonic vacuum pairs, braid all
right members of each pair uniformly to the right and then perform the
quantum walk just on the $n$ strands corresponding to these right members
(see Fig.~\ref{fig:MarkPlatTrace}(a)). Also the reshuffling matrix is taken
to be $\ U=\frac{1}{\sqrt{2}}\left(
\begin{array}{cc}
1 & i \\
i & 1%
\end{array}
\right) $. It is worth noting that the same results reported
below for the $U$ reshuffling matrix are valid if we had employed the
Hadamard matrix $H$, as a straightforward calculation shows.

Next we briefly present the expression for the diagonal part of the walker's
density matrix for the first time steps. For step $t=1,$ we obtain
\begin{equation}
\sum_{\vec{a},\vec{a}^{\prime }}\rho _{\vec{a},\vec{a}^{\prime }}(t=1)|_{%
\text{diag}}=\frac{1}{2}\varphi (B_{2}^{\dagger }B_{2})\left\vert
2\right\rangle \left\langle 2\right\vert +\frac{1}{2}\varphi (B_{3}^{\dagger
}B_{3})\left\vert 4\right\rangle \left\langle 4\right\vert =\frac{1}{2}%
\left\vert 2\right\rangle \left\langle 2\right\vert +\frac{1}{2}\left\vert
4\right\rangle \left\langle 4\right\vert .
\end{equation}%
For step $t=2$ we have
\begin{eqnarray}
\sum_{\vec{a},\vec{a}^{\prime }}\rho _{\vec{a},\vec{a}^{\prime }}(t &=&2)|_{%
\text{diag}}=\frac{1}{4}\Big\{\varphi (\left( B_{1}B_{2}\right) ^{\dagger
}B_{1}B_{2})\left\vert 1\right\rangle \left\langle 1\right\vert +\varphi
(\left( B_{4}B_{3}\right) ^{\dagger }B_{4}B_{3})\left\vert 5\right\rangle
\left\langle 5\right\vert  \notag \\
&&+[\varphi (\left( B_{2}\right) ^{\dagger }B_{2})+\varphi (\left(
B_{3}\right) ^{\dagger }B_{3})]\left\vert 3\right\rangle \left\langle
3\right\vert \Big\}  \notag \\
&=&\frac{1}{4}\left\vert 1\right\rangle \left\langle 1\right\vert
+\frac{1}{2}\left\vert
3\right\rangle \left\langle 3\right\vert +\frac{1}{4%
}\left\vert 5\right\rangle \left\langle 5\right\vert  .
\end{eqnarray}
We notice that the diagonal part of \ the density matrix given above for the
first two steps is common (i.e. independent of $N),$ to all cases of the
symmetric group used i.e. $S_{N}$, $\ N\geq 5.$ This changes for subsequent
steps as we see below.

\bigskip For step $t=3$ we calculate the expression
\begin{eqnarray}
\sum_{\vec{a},\vec{a}^{\prime }}\rho _{\vec{a},\vec{a}^{\prime }}(t &=&3)|_{%
\text{diag}}=\frac{1}{8}\{\left\vert 2\right\rangle \left\langle
2\right\vert +(3+\varphi ((B_{4}^{3})^{\dagger }B_{3}^{2}B_{4})+\varphi
((B_{3}^{2}B_{4})^{\dagger }B_{4}^{3}))\left\vert 4\right\rangle
\left\langle 4\right\vert   \notag \\
&&+(3-\varphi ((B_{6}^{2}B_{5})^{\dagger }B_{5}^{3})-\varphi
((B_{5}^{3})^{\dagger }B_{6}^{2}B_{5}))\left\vert 6\right\rangle
\left\langle 6\right\vert +\left\vert 8\right\rangle \left\langle
8\right\vert \}  \notag \\
&=&\frac{1}{8}\{\left\vert 2\right\rangle \left\langle 2\right\vert
+(3+\varphi (B_{4}^{-2}B_{3}^{2})+\varphi (B_{4}^{2}B_{3}^{-2}))\left\vert
4\right\rangle \left\langle 4\right\vert   \notag \\
&&+(3-\varphi (B_{5}^{2}B_{6}^{-2})-\varphi (B_{5}^{-2}B_{6}^{2}))\left\vert
6\right\rangle \left\langle 6\right\vert +\left\vert 8\right\rangle
\left\langle 8\right\vert \},\text{ \ \ \ \ \ }
\end{eqnarray}%
or in terms of link polynomials \bigskip
\begin{eqnarray}
\sum_{\vec{a},\vec{a}^{\prime }}\rho _{\vec{a},\vec{a}^{\prime }}(t &=&3)|_{%
\text{diag}}=\frac{1}{8}\Big\{\left\vert 2\right\rangle \left\langle
2\right\vert
+(3+d[2,1]^{-9}L(B_{4}^{-2}B_{3}^{2})+d[2,1]^{-9}L(B_{4}^{2}B_{3}^{-2}))%
\left\vert 4\right\rangle \left\langle 4\right\vert   \notag \\
&&+(3-d[2,1]^{-9}L(B_{5}^{2}B_{6}^{-2})-d[2,1]^{-9}L(B_{5}^{-2}B_{6}^{2}))%
\left\vert 6\right\rangle \left\langle 6\right\vert +\left\vert
8\right\rangle \left\langle 8\right\vert \Big\}  \notag \\
&=&\frac{1}{8}\Big\{\left\vert 2\right\rangle \left\langle 2\right\vert
+(3+2d[2,1]^{-2}\big[ 1+\frac{1}{2}(N-2)(N-3)\big] ^{2})\left\vert
4\right\rangle \left\langle 4\right\vert   \notag \\
&&+(3-2d[2,1]^{-2}\big[ 1+\frac{1}{2}(N-2)(N-3)\big] ^{2})\left\vert
6\right\rangle \left\langle 6\right\vert +\left\vert 8\right\rangle
\left\langle 8\right\vert \Big\},
\end{eqnarray}%
where (\ref{lin}) has been used. By means of (\ref{link}) we obtain
\begin{eqnarray*}
L(B_{5}^{2}B_{6}^{-2})
=L(B_{5}^{-2}B_{6}^{2})=L(B_{4}^{-2}B_{3}^{2})=L(B_{4}^{2}B_{3}^{-2}) =\big[
1+\frac{1}{2}(N-2)(N-3)\big] ^{2}d[2,1]^{7}
\end{eqnarray*}
For the choice $N=5,$ the dimension is $d[2,1]=10,$ and this yields
\begin{equation}
\rho (t=3)|_{diag}=\frac{1}{8}\Big\{\left\vert 2\right\rangle \left\langle
2\right\vert +(3+\allowbreak \frac{8}{25})\left\vert 4\right\rangle
\left\langle 4\right\vert +(3-\allowbreak \frac{8}{25})\left\vert
6\right\rangle \left\langle 6\right\vert +\left\vert 8\right\rangle
\left\langle 8\right\vert \Big\}.
\end{equation}

Similarly for the $t=4$ step we obtain
\begin{eqnarray}
\sum_{\vec{a},\vec{a}^{\prime }}\rho _{\vec{a},\vec{a}^{\prime }}(t &=&4)|_{%
\text{diag}}=\frac{1}{16}\left\vert 1\right\rangle \left\langle 1\right\vert
+\Big\{\frac{1}{4}+\frac{3}{8}\big[ 1+\frac{1}{2}(N-2)(N-3)\big]
^{2}d[2,1]^{-2}\Big\}\left\vert 3\right\rangle \left\langle 3\right\vert
\notag
\\
&&+\Big\{\frac{3}{8}-\frac{1}{4}\big[ 1+\frac{1}{2}(N-2)(N-3)\big]
^{2}d[2,1]^{-2}\Big\}\left\vert 5\right\rangle \left\langle 5\right\vert
\notag
\\
&&+\Big\{\frac{4}{16}-\frac{2}{16}\big[ 1+\frac{1}{2}(N-2)(N-3)\big]
^{2}d[2,1]^{-2}\Big\}\left\vert 7\right\rangle \left\langle 7\right\vert
+\frac{1}{16}\left\vert 9\right\rangle \left\langle 9\right\vert ,
\end{eqnarray}%
which for the $N=5$ and $d[2,1]=10$ give the diagonal part
\begin{equation}
\sum_{\vec{a},\vec{a}^{\prime }}\rho _{\vec{a},\vec{a}^{\prime }}(t=4)|_{%
\text{diag}}=\frac{1}{16}\left\vert 1\right\rangle \left\langle 1\right\vert
+\frac{31}{100}\left\vert 3\right\rangle \left\langle 3\right\vert +\frac{67%
}{200}\left\vert 5\right\rangle \left\langle 5\right\vert +\frac{23}{100}%
\left\vert 7\right\rangle \left\langle 7\right\vert +\frac{1}{16}\left\vert
9\right\rangle \left\langle 9\right\vert .
\end{equation}

\begin{figure}[h]
\begin{center}
\includegraphics[scale=0.5]{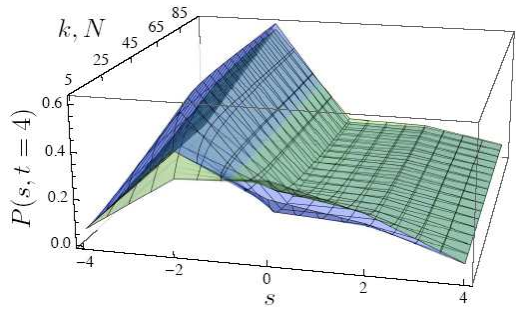}
\caption{\label{fig:newfig} Combined surface plot for the distribution of a $t=4$ walk in
the modified protocol that corresponds to the Markov trace. The blue surface
is the distribution for spin-$1/2$ anyons in the quantum group $SU(2)_k$ and
the green surface for quantum double anyons in the $(2,1)$ irrep of
$D(S_N)$.}
\end{center}
\end{figure}

Before closing this section some comments are in order. Comparison of the
diagonal parts of walkers's density matrix obtained above for the two forms
of modelling the braid group representations, i.e. quantum group $SU(2)_{k}$
and symmetric group respectively, shows that the walker occupies the same
sites at both cases albeit with different occupation probabilities. This
difference in the distribution of \ occupation probabilities is expected to
be a general feature, since by construction these probabilities depend both
on the algebraic structure (finite group, quantum group etc.), employed to
provide a matrix representation of the braid group, via the associated link
polynomials. It is worth mentioning that in the limits $k\rightarrow
\infty$ and $N\rightarrow \infty$, both distributions converge to the standard
quantum walk as seen in Fig.~\ref{fig:newfig}.

\section{Outlook}
\label{outlook}

We have described here a simple quantum walk on a line segment with $n $
sites. This is a path graph with vertices given by the locations of $n$
anyons and $n-1$ edges connecting the vertices that correspond to generators
of the braid group $B_n$.  One could imagine anyonic walks on more general
graphs by embedding the graph onto a cellular surface of genus $g$.  Such a
mapping is not unique and the study of such embeddings is the focus of
topological graph theory~\cite{GT}. One could then perform walks with an
anyon on this higher genus surface using the representation theory for braid
groups on arbitrary manifolds~\cite{Birman}.

There are numerous questions on anyonic quantum walks that have not been
addressed yet. The Abelian anyonic quantum walk is easily tractable, both
theoretically and numerically, giving rise to a distribution similar to the
standard quantum walk. As we have seen one could determine the behavior of
non-Abelian quantum walks for a small number of steps. It is a very hard
problem to probe any non-Abelian walk for large anyonic chains due to the
exponential increase in their Hilbert space dimension. Each additional step
of the walker exponentially increases the number of different paths that one
needs to consider. On the top of that one needs to evaluate the Kauffman
brackets of the corresponding paths in order to determine the final position
distribution of the walker. For different types of anyonic models (like
SU$(2)_2$) the evaluation of the Jones polynomials is computationally
polynomially demanding~\cite{Lickorish,Jaeger}, though still one has to
consider the exponential increase in the number of paths. For other models
(like SU$(2)_k$ for $k>2$ and $k\neq 4$) the Jones polynomial evaluation is
believed to be, in general, exponentially hard to compute~\cite{Jaeger,Kuperberg}.

Ideally, one is interested in deriving the asymptotic behavior of
non-Abelian anyonic quantum walks. The examples we considered here reveal
parts of the structure and possibly indicate the asymptotic behavior of
certain models. For example, we explicitly considered the SU$(2)_k$ anyonic
model for up to $t=10$ number of steps. This gave the following interesting
characteristics. For small $k$ the walk appeared to have a distribution
close to the classical random walk. For large $k$ the distribution appeared
to be similar to a quantum. Even though small $t$ distributions can hardly
be conclusive about the large $t$ behavior they provide the exciting
possibility of having an anyonic model with intermediate values of $k$ where
the variance is in between the classical and the quantum. That would constitute a new paradigm of diffusive behavior. A variety of non-Abelian models needs
to be systematically considered theoretically and numerically in order to
conclusively understand the asymptotic behavior of non-Abelian walks.

\begin{acknowledgements}

This work was supported by the EU grants EMALI and SCALA, UK Engineering and Physical Sciences Research Council and the Royal Society.

\end{acknowledgements}

\end{document}